\documentclass{article}

\usepackage[final,nonatbib]{nips_2018}
\usepackage{graphicx}

\usepackage{graphicx}
\usepackage{subfigure}
\graphicspath{ {./images/} }
\usepackage[utf8]{inputenc} 
\usepackage[T1]{fontenc}    
\usepackage{hyperref}       
\usepackage{url}            
\usepackage{booktabs}       
\usepackage{amsfonts}       
\usepackage{nicefrac}       
\usepackage{microtype}      
\usepackage{amsmath}
\usepackage[sorting=none]{biblatex}
\title{Comparative Analysis of Vision Transformers and Traditional Deep Learning Approaches for Automated Pneumonia Detection in Chest X-Rays}
\author{
   Gaurav Singh \\
   \texttt{grvsingh@g.ucla.edu}
}

\begin{document}

\maketitle
\begin{abstract}
Pneumonia, particularly when induced by diseases like COVID-19, remains a critical global health 
challenge requiring rapid and accurate diagnosis. This study presents a comprehensive comparison 
of traditional machine learning and state-of-the-art deep learning approaches for automated 
pneumonia detection using chest X-rays (CXRs). We evaluate multiple methodologies, ranging from 
conventional machine learning techniques (PCA-based clustering, Logistic Regression, and Support 
Vector Classification) to advanced deep learning architectures including Convolutional Neural 
Networks (Modified LeNet, DenseNet-121) and various Vision Transformer (ViT) implementations 
(Deep-ViT, Compact Convolutional Transformer, and Cross-ViT). Using a dataset of 5,856 
pediatric CXR images, we demonstrate that Vision Transformers, particularly the Cross-ViT 
architecture, achieve superior performance with 88.25\% accuracy and 99.42\% recall, surpassing 
traditional CNN approaches. Our analysis reveals that architectural choices impact performance 
more significantly than model size, with Cross-ViT's 75M parameters outperforming larger 
models. The study also addresses practical considerations including computational efficiency, 
training requirements, and the critical balance between precision and recall in medical 
diagnostics. Our findings suggest that Vision Transformers offer a promising direction for 
automated pneumonia detection, potentially enabling more rapid and accurate diagnosis during 
health crises.
\end{abstract}

\text{*The implementation code for this research can be found here: \cite{repolink}}
\section{Introduction}
\begin{figure}[h!]
  \centering
  \includegraphics[scale=0.35]{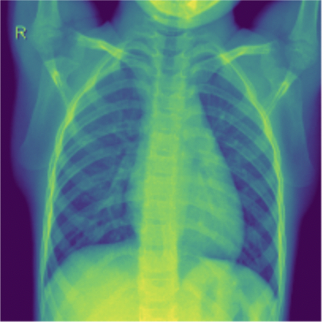}
  \hspace{0.5cm}
  \includegraphics[scale=0.35]{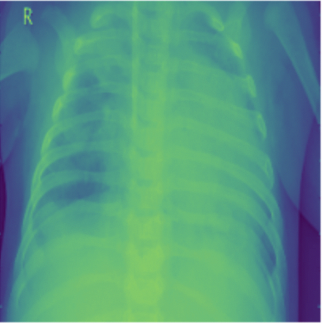}
  \caption{Normal vs Pneumonia infected CXR}
  \label{comp}
\end{figure}

Pneumonia is an acute pulmonary infection which causes inflammation of air sacs called Alveoli. It is most common in underdeveloped or developing nations which have below average air quality and sanitation standards. Pneumonia can be of these three forms viral, bacterial or fungal which makes it airborne and contagious. It mostly effects children under the age of five or elderly people who have weak immune system \cite{who}. Symptoms include coughing, fever, difficulty in breathing, loss of appetite etc. and is life threatening. COVID-19 induced Pneumonia has been a major reason for pandemic deaths in the world \cite{stats}. Pneumonia is a rapid attacking infection whose quick diagnosis is important for saving lives. 
Detection of Pneumonia normally requires a highly skilled doctor to look at the Chest X-rays (CXR) and CT scans and requires medical history to arrive at a conclusion. Normally a foggy CXR with white spots are a few indicators of Pneumonia Fig.~\ref{comp}. This manual process might be expensive and critically time consuming. The recent advancements in Machine learning and Computer vision might prove really helpful in quick and accurate detection of Pneumonia and can potentially save a lot more lives especially during pandemics like COVID-19 which faced shortage of human expertise.

\section{Data}
\begin{figure}[h!]
  \centering
  \includegraphics[scale=0.30]{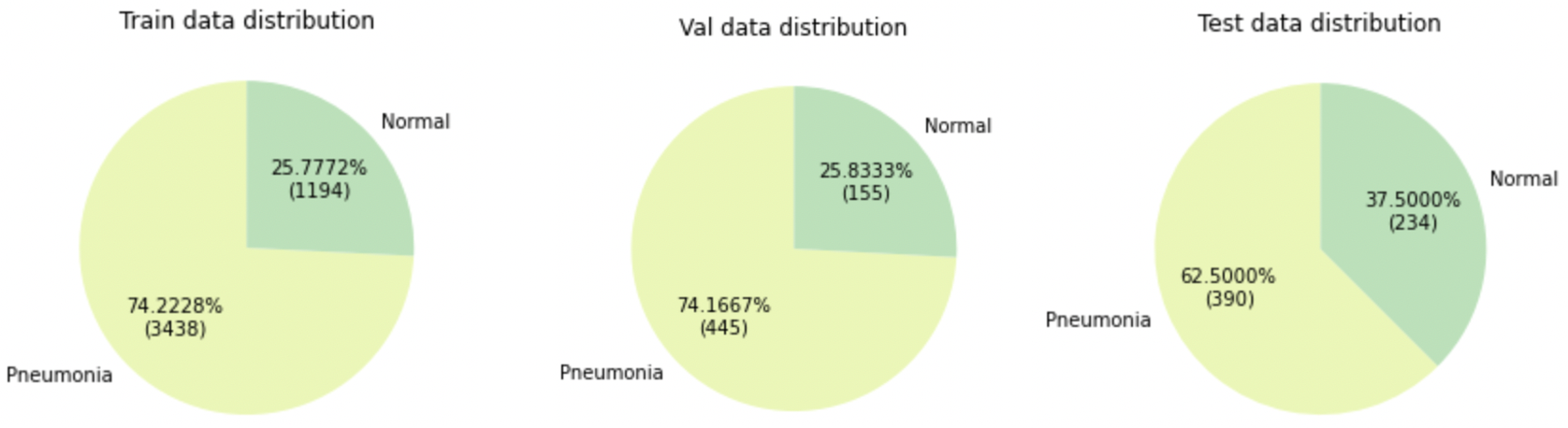}
  \caption{Class Distribution}
  \label{dist}
\end{figure}

The data used for this project is based on data originally provided by Kermany et al \cite{odata} which is available as open source on Kaggle \cite{data}. There are a total of 5,856 CXR(anterior-posterior) images from pediatric patients of one to five years old. The raw data is organized into three - train, test and validation - folders. Each folder is further having images into two sub-folders namely Pneumonia and Normal. There are 5,216 train, 624 test and 16 validation images originally which I redistributed using Stratified sampling so as to get 600 images for both train and test data which is around 10\% of total data. Class distribution of data is shown in Fig.~\ref{dist} which tells that there is a heavy class imbalance with $\sim75\%$ images belonging to positive patients. The images are of varying dimensions therefore all images are re-scaled to 256 X 256 while read. It is noticed that there is channel inconsistency among images with 251 images being RGB whereas 4,381 images being gray-scale. All images therefore are read as gray-scale to maintain consistency.

\section{Unsupervised Machine Learning}
\begin{figure}[h!]
  \centering
  \includegraphics[scale=0.35]{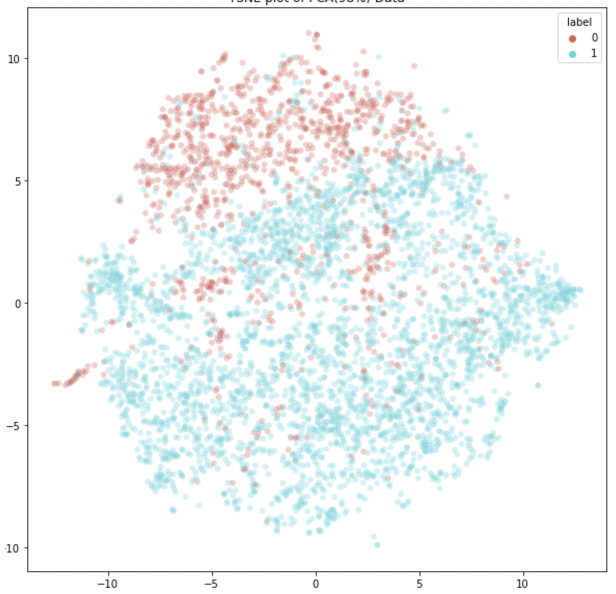}
  \caption{t-SNE plot of train images}
  \label{cluster}
\end{figure}

For the start, I investigated the two classed and is their any significant difference in the data distribution that can be segregated into two groups. For this a relevant work is already done for Brain tumor segmentation using Principal components based clustering \cite{kaya2017pca}. I took inspiration from this paper and used SVD decomposition instead of EM based PCA to find principal components which retain 98\% variance. Each train image is unrolled into a 65,536 dimension row vector and I kept 1,648 principal components out of it for faster computation. K-means clustering with 2 clusters, gave two centroids which are then used to classify the test images based on Euclidean distance from cluster centroids. The train imageset is visualised in 2 dimensional space using t-SNE components instead of PCA since it's a non-linear transformation unlike PCA which preserve the Local structure of data by minimizing the Kullback–Leibler divergence (KL divergence). The visualisation Fig.~\ref{cluster} of the train data shows that there is a significant difference between the two class images and thus can be segregated using a good decision boundary.

\section{Supervised Machine Learning}
\subsection{Logistic Regression}
Since Pneumonia detection is a binary classification task Logistic regression is a good baseline to start with. It is just like linear regression with values mapped to probabilities using Sigmoid function and it uses a log loss or binary cross entropy loss function for training. For this approach, Principal components of images are used instead of full length data. Regularization value of 0.001 is used to avoid over-fitting and class weights are balanced by a factor of inverse of no. of per class samples to deal with imbalanced dataset.

\subsection{Support Vector Classifier}
Logistic regression decision boundary is linear and it's shown in Fig.~\ref{cluster} that a non-linear decision will be a good choice. I used Soft margin Support Vector Classifier (C = 1) with Radial Basis Function (RBF) as kernel to get non-linear decision boundary. Principal components are used instead of full data and the weights are balanced here as well.

\section{Convolution Networks}
Convolution networks provided the major shift in the computer vision tasks by using the concept of filters which go through the full image and generate feature maps using element wise matrix multiplications. They have less number of parameters to work with given filters have less trainable parameters and full connections are limited to last few layers.
\subsection{LeNet architecture}
This is the first CNN architectue which uses 4 [CONV-POOL] layer and 3 Fully connected(FC) layers as shown in Fig.~\ref{cnns}(a). The network is modified a bit with Batchnorm and ReLU as activation function. Batchnorm helps to manage Intern Covariate Shift which enables us to use higher learning rate and has regularization effect.

\subsection{DenseNet-121 architecture}
\begin{figure}[h!]
    \centering
    \subfigure[]{\includegraphics[scale=0.15]{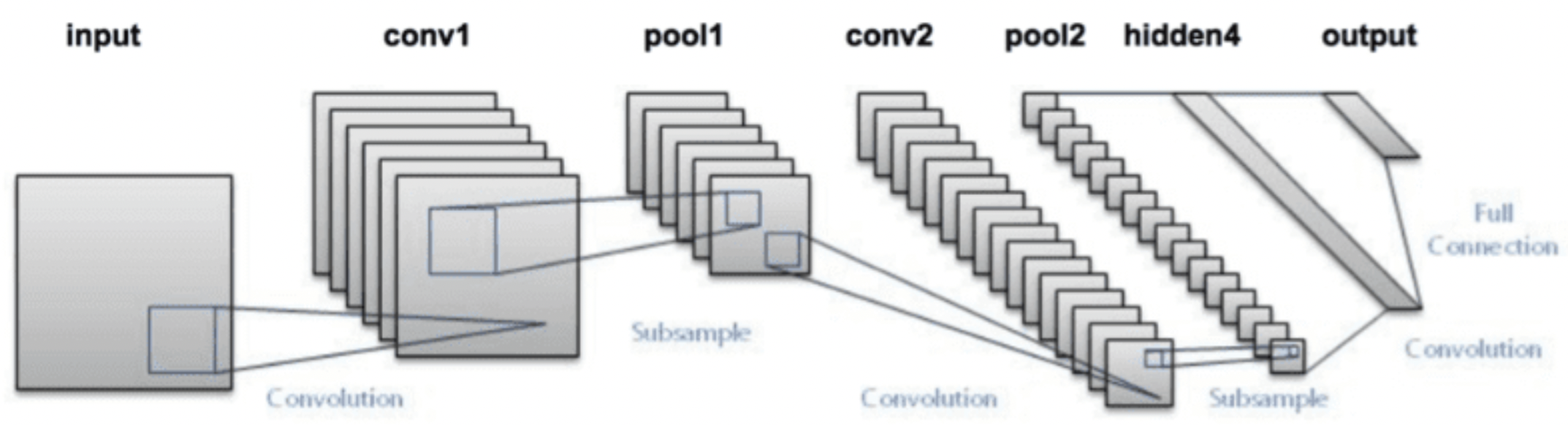}} 
    \hspace{0.5cm}
    \subfigure[]{\includegraphics[width=6cm, height=1.5cm]{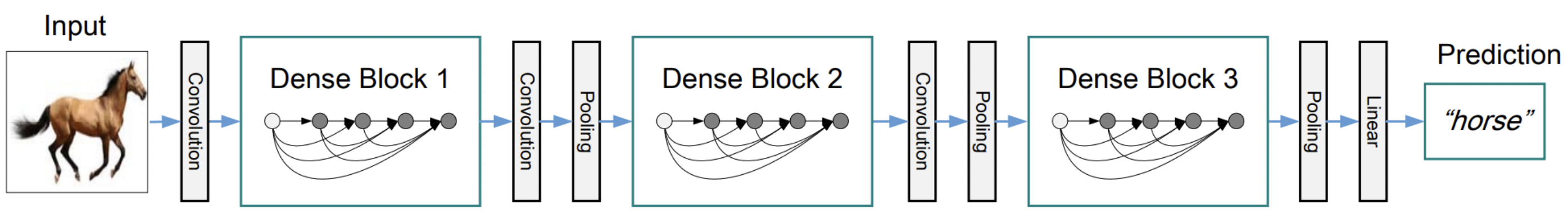}} 
    \caption{(a) LeNet-5 (b) DenseNet-121}
    \label{cnns}
\end{figure}

DenseNet-121 is a 121 layered CNN which was used by Rajpurkar et al. \cite{rajpurkar2017chexnet} for pneumonia classification. DenseNet uses dense blocks as shown in Fig.~\ref{cnns}(b) to pass all previous layers outputs to current layer to propagate information to great depths and solves the problem of vanishing gradients. To keep the dimensions in control from all these previous output concatenations, 1x1 convolutions in Transition layers are used between each Dense Block.

\vspace{-0.25cm}
\section{Vision Transformers (ViT)}
\vspace{-0.25cm}
CNNs are gradually being replaced by the State-of-the-art Vision Transformers proposed by Dosovitskiy et al \cite{dosovitskiy2020image} because of their overall computation efficiency, easy GPU parallelisation, ViTs don’t require large number of layers as CNN for comparable feature representations and have greater receptive field. They incorporate skip connections in each encoder and decoder and thus are not susceptible to the problem of vanishing gradients. Transformers are used in Natural Language Processing(NLP) tasks but if we break the image into patches which are analogous to words and flatten these patches which are analogous to word embeddings and feed it to a transformer we get a vision transformer.

\vspace{-0.25cm}
\subsection{Deep Vision Transformer}
\vspace{-0.25cm}
Deep ViT proposed by Zhou et al \cite{zhou2021deepvit} as shown in Fig.~\ref{vits}(a) is just like a basic ViT but with increased number of encoder blocks. I have used 12 such encoder blocks with 24 multi attention heads to compute more feature maps. The patch size used is 32.

\subsection{Compact Convolution Transformer (CCT)}
\vspace{-0.25cm}
CCT proposed by Hassani et al \cite{hassani2021escaping} as shown in Fig.~\ref{vits}(b) uses compact transformers by using convolutions instead of patching and performing sequence pooling for reducing trainable parameters. This allows for CCT to have high accuracy and a low number of parameters. Also using a mix of both convolution feature maps and simple patches as input reduce the inductive bias and helps transformer to learn better representations. I used 6 convolution layers with kernel size as 3 x 3 to get intitial feature maps and 6 multi attention heads and 14 encoder layers.

\subsection{Cross Vision Transformer}
\begin{figure}[h!]
    \centering
    \subfigure[]{\includegraphics[scale=0.20]{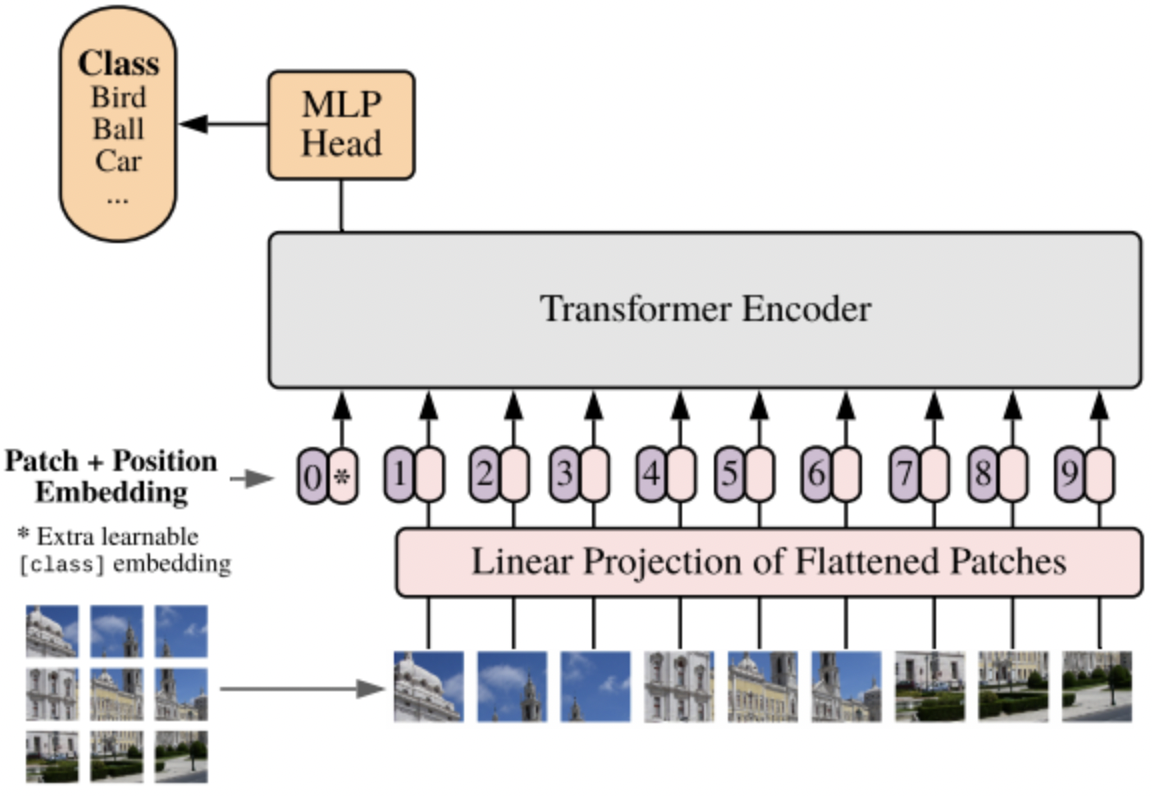}} 
    \hspace{0.5cm}
    \subfigure[]{\includegraphics[scale=0.10]{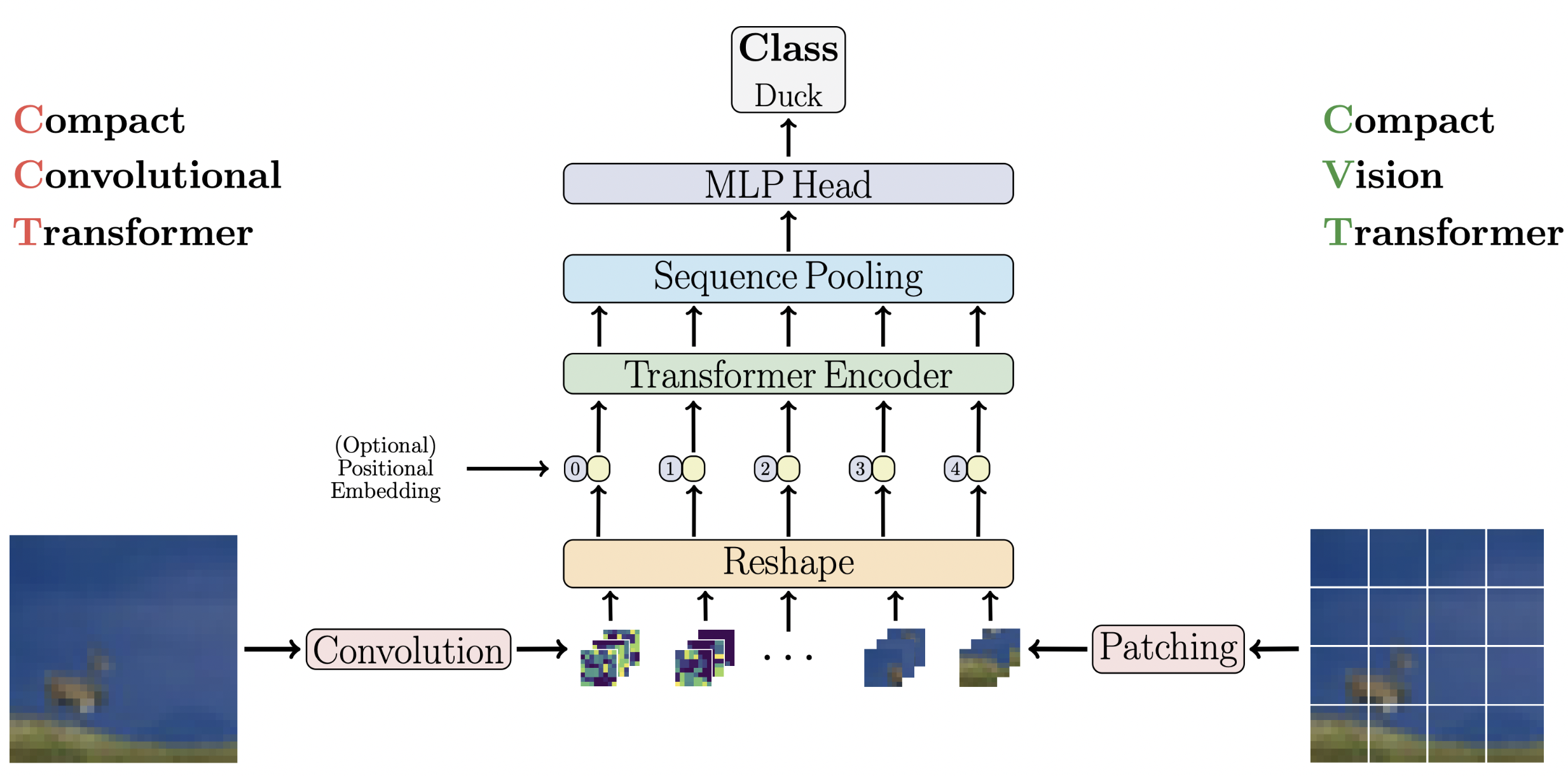}}
    \hspace{0.5cm}
    \subfigure[]{\includegraphics[scale=0.20]{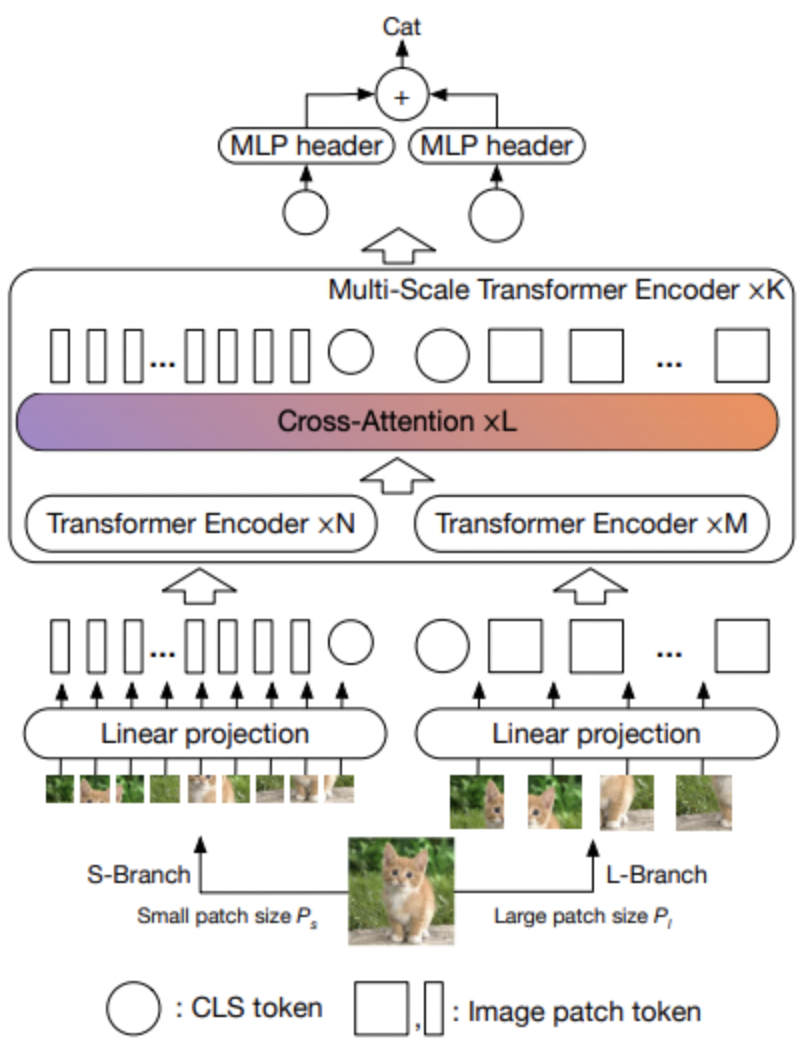}}
    \caption{(a) Deep ViT (b) CCT (c) Cross ViT}
    \label{vits}
\end{figure}

\vspace{-0.25cm}
Chen et al \cite{chen2021crossvit} proposes to have two vision transformers processing the image at different scales as shown in Fig.~\ref{vits}(c), cross attending to one every so often. They show improvements on top of the base vision transformer. I have used the upscaled dimensions as 384 x 384 and downscaled dimensions as 192 x 192 with 8 multi attention heads and 3 encoder layers.

\section{Hyperparameters, Initialisation and Training strategy}
Unsupervised methods and supervised machine learning approaches are trained on a 16 core CPU machine whereas the Deep learning approaches i.e. CNNs and ViTs are trained and tested on 6 Nvidia Tesla K80 GPU using the data parallelism approach for fast, efficient and distributed training.
For Machine learning approaches i.e. Logistic regression and SVC best hyperparameters (kernel type, regularization coeff.) are found using grid search methods.

For Deep Learning approaches, Xavier initialisation is used for fully connected layers and Kaiming initialisation is used for Convolution layers. For DenseNet-121, the initial weights are taken from model pre-trained on Imagenet dataset. Loss function used during training is cross entropy to which the softmax outputs are fed. Unsupervised, Machine learning and LeNet architecture is trained on single channel images whereas Deep Learning(DL) architectures are trained using 3-channel images formed by stacking the same grayscale image three times. 

For DL methods, each image was rescaled to a dimension of 224 X 224 from 256 X 256 and since the training data is less and given DL training is data intensive, each image is horizontally flipped, scaled and shifted (by a factor of 0.1) and rotated (by a factor of 5 degrees) with a probability of 0.5 to augment the training data. Data augmentation helps in better learning and helps in avoiding over-fitting. Each image is then normalised by using the mean = 0.48 and standard deviation = 0.22 which are found from the train images. Different learning rate scheduling methods are used with different DL approach. For ViT, 10\% of total steps are used as warm-up steps with learning rate of 1e-4 and linearly decayed scheduler is used. For DenseNet-121, cosine decay scheduler is used with learning rate of 1e-4. It is noted that transformer models don't train well i.e. loss doesn't decrease, if warm-up steps are not used. Adam optimizer is used for training.

\section{Results discussion and Conclusion}

\begin{table}[h!]
\centering
 \begin{tabular}{||c| c| c| c| c| c||} 
 \hline\hline
 Model & Acc\% & Precsion & Recall & Train Time(s) & Test Time(s) \\ [0.5ex] 
 \hline
 PCA + Clustering & 58.17 & 68.48 & 61.28 & 279.53 & 2.21 \\ 
 Logistic Regression & 71.16 & 68.04 & 94.46 & 105.65 & 0.035 \\
 SVC & 75.00 & 71.64 & 93.94 & 5.99 & 0.803 \\
 \hline
 \end{tabular} \\ [1ex] 
 \caption{Results: ML Approaches}
 \label{mltable}
\end{table}

\begin{table}[h!]
\centering
 \begin{tabular}{||c| c| c| c| c| c | c||} 
 \hline\hline
 Model & Acc\% & Precsion & Recall & Epochs & Train Time(s) & Test Time(s) \\ [0.5ex]
 \hline
 Mod. LeNet & 77.40 & 73.44 & 1.0 & 50 & 150 & 3 \\ 
 DneseNet-121 & 83.05 & 79.80 & 99.74 & 50 & 4000 & 16 \\
 Deep-ViT & 80.60 & 78.01 & 99.23 & 20 & 3060 & 13 \\
 CCT & 84.08 & 81.17 & 1.0 & 20 & 1200 & 7 \\
 Cross-ViT & 88.25 & 84.02 & 99.42 & 20 & 3780 & 18 \\
 \hline
 \end{tabular} \\ [1ex] 
 \caption{Results: DL Approaches}
 \label{dltable}
\end{table}

\begin{figure}[h!]
    \centering
    \subfigure[]{\includegraphics[scale=0.35]{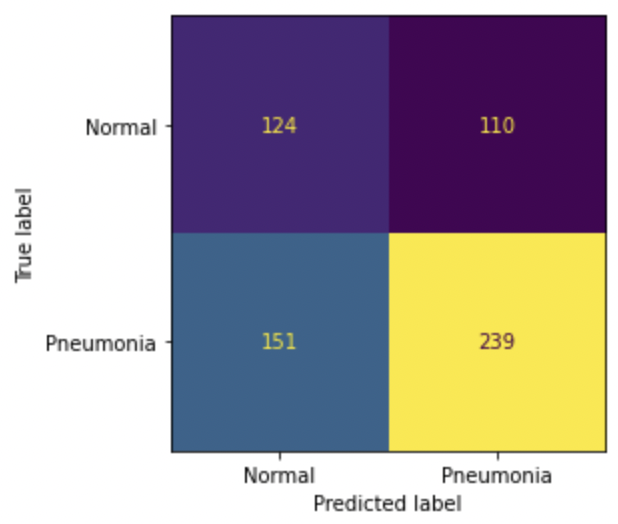}} 
    \hspace{0.5cm}
    \subfigure[]{\includegraphics[scale=0.35]{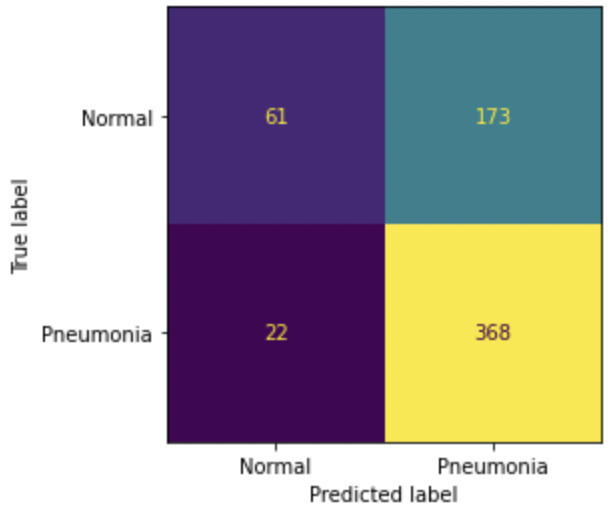}}
    \hspace{0.5cm}
    \subfigure[]{\includegraphics[scale=0.35]{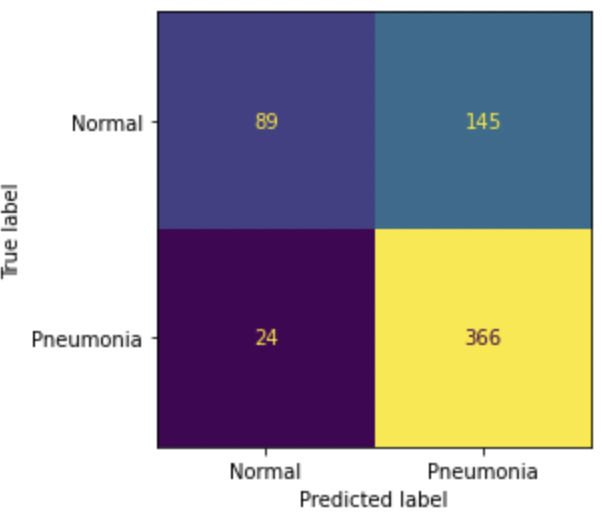}}
    \caption{(a) PCA+Clustering (b) Logistic Regression (c) SVC}
    \label{confml}
\end{figure}

From Table \ref{mltable}, we can see that the ML approaches didn't do well in terms of our evaluation metrics i.e Accuracy, precision and recall. For medical diagnosis, high Recall is much more important i.e. to minimize the False Negatives, also a high Precision is desired, given too many False Positives can lead to wastage of time in critical times like COVID-19. PCA + Clustering gave poorest results and took the most amount of train and test time due to PCA decomposiion. Logistic regression and SVC did well with a high recall but still >20 False negatives as seen from Fig. \ref{confml}. One thing to note is that SVC can't be used with large no. of Principal components or with large no. of training samples given its runtime is cubic in the max(\# samples, \#features). One reason that ML approaches are not able to do well is that by flattening an image we loose spatial structuring which is important for any image identification. Also the decision boundary of LR is linear whereas for SVC is not that complex to segregate the subtle image variations.

\begin{figure}[h!]
    \centering
    \subfigure[]{\includegraphics[scale=0.35]{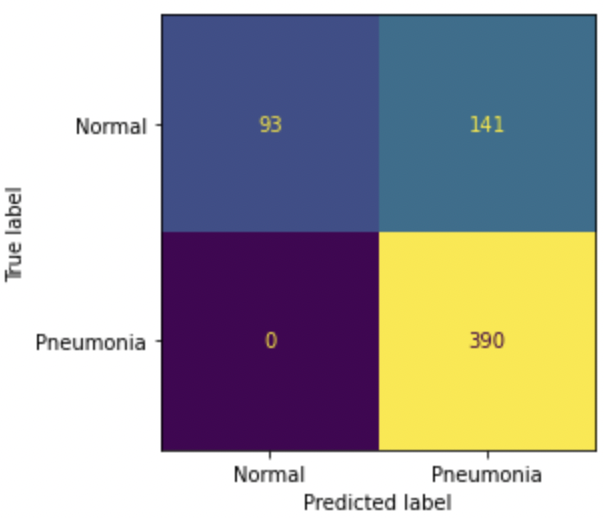}} 
    \hspace{0.5cm}
    \subfigure[]{\includegraphics[scale=0.35]{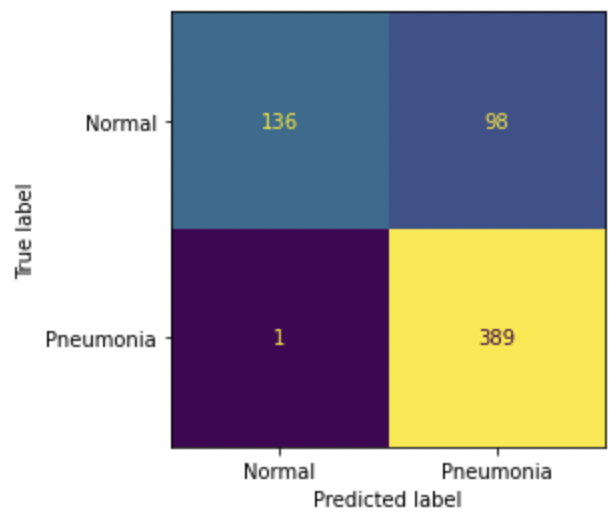}}
    \hspace{0.5cm}
    \subfigure[]{\includegraphics[width=5cm, height=3.5cm]{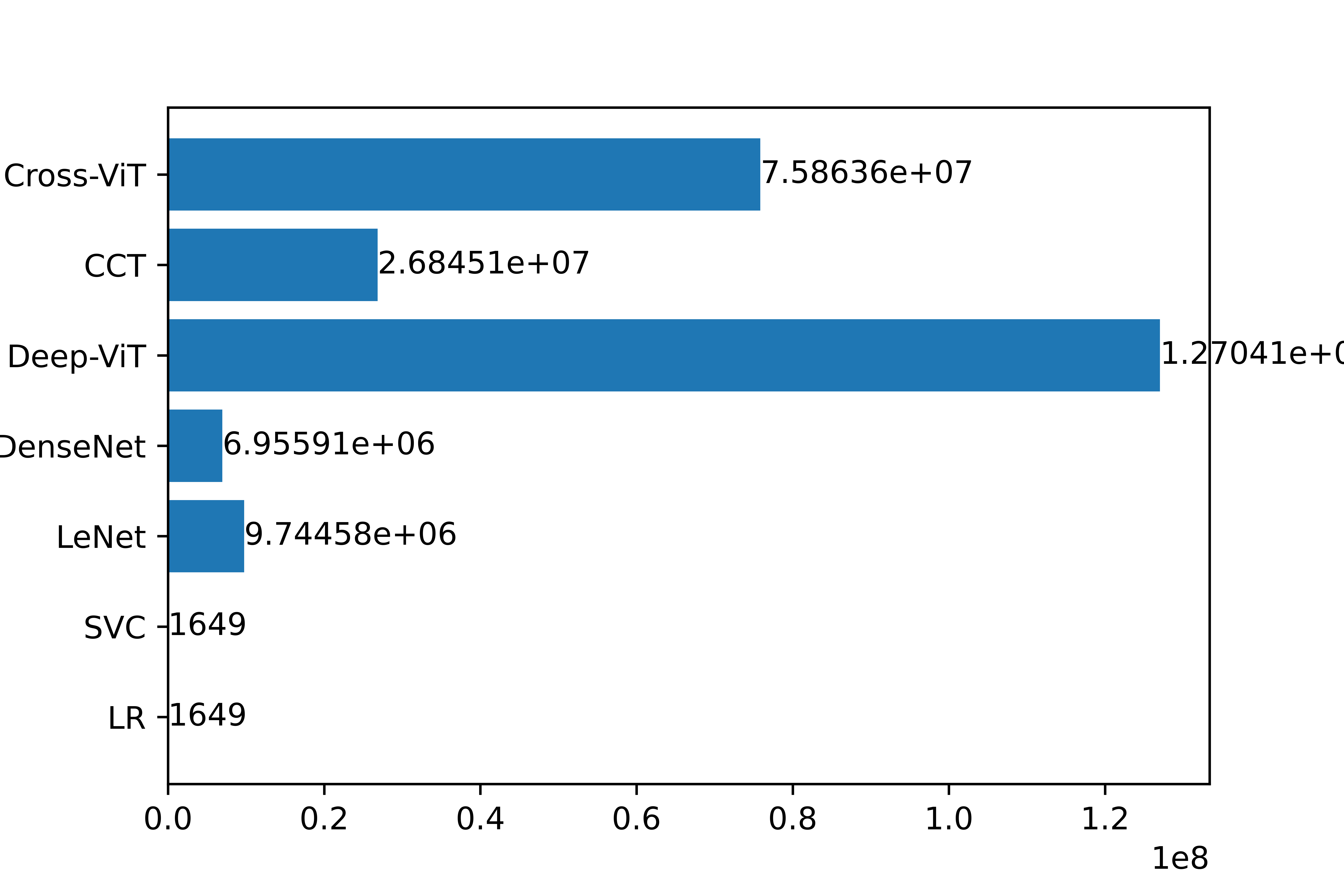}}
    \caption{(a) Modified LeNet-5 (b) DenseNet-121 (c) No. of trainable parameters}
    \label{confcnn}
\end{figure}

\begin{figure}[h!]
    \centering
    \subfigure[]{\includegraphics[scale=0.35]{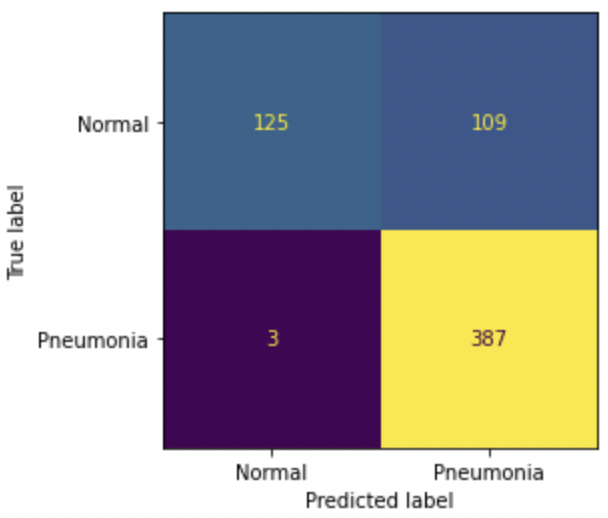}} 
    \hspace{0.5cm}
    \subfigure[]{\includegraphics[scale=0.35]{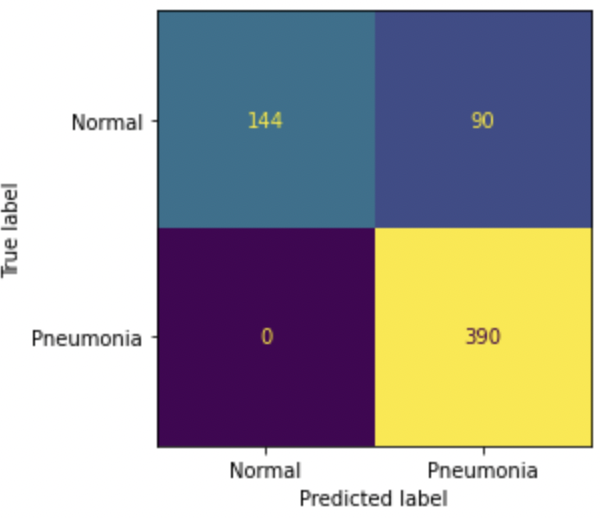}}
    \hspace{0.5cm}
    \subfigure[]{\includegraphics[scale=0.35]{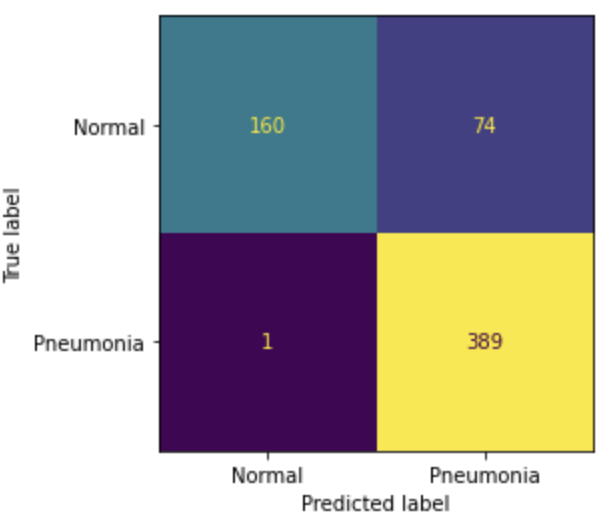}}
    \caption{(a) Deep-ViT (b) CCT (c) Cross-ViT}
    \label{confdl}
\end{figure}

From Table \ref{dltable} we can see that ViTs did well as compared to CNN models in terms of all metrics giving high precision and recall, with less no. of epochs to get the same or even better results. DenseNet performed well but it has a large number of layers and it uses pretrained weights from Imagenet which are not available off the shelf for ViTs yet.
A standout metric for DL methods is that, recall is very high which is very much desired as seen from Fig. \ref{confcnn} and Fig. \ref{confdl}. Among CNNs, DenseNet achieved best results by far with 83.05\%accuracy and a very high recall of 99.74\%. It would be worth to try more complex architectures like ResNet-50 or InceptionV2. 

It's quite fascinating how well ViTs performed even without any pre-training and given that Transformer models only start to do well after they are trained for large no. of epochs on large relevant datasets. Cross-ViT achieved the best results \ref{dltable} with 88.25\% accuracy and 99.42\% recall. The next step can be to pretrain the ViTs on Imagenet dataset and then to fine tune the model on Pneumonia dataset.
From Fig. \ref{confcnn}(c) we can see that Deep-ViT has around 127M parameters but the results are not good when compared to Cross-ViT having only 75M parameters. Similar thing can be seen between LeNet(9.7M) and DenseNet(6.9M). Thus, having a large no. of parameters will not guarantee towards better results, the choice of DL architecture is an important aspect. From Table \ref{dltable}, per epoch train time is large for ViTs given they have on avg. around >70 M parameters to train. 

On analysing, the results I found that we are not doing well to mitigate False positives, one reason for that could be class imbalance given pneumonia positive samples are large. Also we can do much better if we use pre-trained models and just train the fully connected layers instead of training full models on such a small dataset. From Table \ref{dltable} and Table \ref{mltable}, the test times are within reasonable limits making ML/ DL techniques good for faster and accurate diagnosis of diseases on scale in the future.

\printbibliography
\end{document}